\documentclass[12pt]{article}
\usepackage[utf8]{inputenc}
\usepackage[english]{babel}
\usepackage{amsfonts}
\usepackage{textcomp}
\usepackage{amsmath, amssymb, graphicx}
\renewcommand{\theequation}{\arabic{section}.\arabic{equation}}
\usepackage{geometry}
\usepackage{enumerate}
\usepackage{hyperref}
\usepackage{braket}
\usepackage{color}
\usepackage{cite}
\usepackage{float}
\usepackage{indentfirst}
\usepackage{epsfig}
\usepackage{booktabs}
\usepackage{subfigure}
\usepackage{caption}
\usepackage{ulem}
\usepackage{cancel,soul}
\usepackage{array}
\usepackage{multirow}

\newcommand{\be}{\begin{equation}}
\newcommand{\ee}{\end{equation}}
\newcommand{\bea}{\begin{eqnarray}}
\newcommand{\eea}{\end{eqnarray}}
\newcommand{\nn}{\nonumber}
\newcommand{\lb}{\label}

\newcommand{\subsec}[1]{%
  \par\noindent
{\normalsize\bfseries\itshape #1}%
  \enspace
}

\title{Wigner continuous-spin equations in $\mathbf{AdS_D}$: \\ bosonic and fermionic  cases} 

\author{
  \small{{\bf Lev N. Astrakhantsev,}$^{a,b}$}\footnote{\tt levastr@theor.jinr.ru} \,
  \small{{\bf Anastasia A. Golubtsova}$^{a}$}\footnote{\tt golubtsova@theor.jinr.ru} \, \\
  \small{{\bf and Mikhail A. Podoinitsyn}$^{a}$}\footnote{\tt mpod@theor.jinr.ru}
}
\date{\begin{small}
  $^{a}$ Bogoliubov Laboratory of Theoretical Physics, JINR, \\
  141980 Dubna, Moscow Region, Russia \\
  $^{b}$ Moscow Institute of Physics and Technology, \\
  Institutskii per, 9, Dolgoprudny, 141700, Russia
  \end{small}
}

\begin{document}

\maketitle

\begin{abstract}
We construct Wigner-like equations of motion for symmetric continuous-spin fields in anti-de Sitter space of arbitrary dimension, treating both the bosonic and fermionic cases. We generalize the classical flat-space Wigner constraints for bosonic continuous-spin fields, and for the fermionic case we adopt the equations proposed by Bekaert and Mourad as our starting point. This is achieved by covariantising the ordinary derivatives and deforming the resulting constraints so that they form a closed algebra. The construction is carried out in a metric-like formalism and yields a system of first-class constraints that define a representation of the $\mathfrak{so}(2,D-1)$ isometry algebra, realized via the Lie-Lorentz derivative. Using these constraints, we compute the eigenvalues of the quadratic and quartic Casimir operators and compare the obtained values with Metsaev's classification of continuous-spin representations for both the bosonic and fermionic cases. A crucial algebraic role is played by special operators of the (super)algebra Howe-dual to the Lorentz subalgebra $\mathfrak{so}(1,D-1)$: the $\mathfrak{sl}(2)$ Casimir operator in the bosonic case, and the Casimir's ghost of the $\mathfrak{osp}(1|2)$ superalgebra in the fermionic case. Both objects naturally organize the constraint algebra and ensure its consistency.
\end{abstract}

\newpage

\tableofcontents

\newpage

\section*{Introduction}
\addcontentsline{toc}{section}{Introduction}

Continuous-spin (CS) fields represent physical systems with an infinite number of spin degrees of freedom, characterized by a continuous dimensionful parameter $\boldsymbol{\mu}$ (with dimension of inverse length), and correspond to unitary irreducible representations (UIRs) of the $D$-dimensional Poincar\'e group \footnote{In fact, when we say that the UIR of CS is characterized by a single continuous parameter $\boldsymbol{\mu}$, we are referring only to the massless CS. Massive UIRs of CS also exist \cite{BeBo}, albeit only for purely imaginary mass. In this work, we restrict ourselves to the massless CS, but the massive case is also actively discussed in the literature, and some of the works cited in our introduction treat it on an equal footing with the massless one.}.
In four-dimensional Minkowski spacetime, the continuous-spin representation is realized on fields satisfying the Wigner equations \cite{w-1, wb}.  Although questions regarding the physical properties of such fields were raised early on \cite{w-2, ww-2, w-3}, their unusual characteristics proved challenging to interpret \footnote{It was later shown that continuous-spin representations can be consistently incorporated into the framework of local quantum theory \cite{l-8}.}, causing interest to wane for a period. Research on continuous-spin fields has gained fresh momentum over the past two decades, 
 largely stimulated by breakthroughs in higher-spin theories \cite{MV-1, MV-2, reviewsV, bengtsson2020vol1, bengtsson2020vol2} and potential connections to string theory \cite{Savvidy:2004}. In particular, the work \cite{l-1-1} provided a systematic $D$-dimensional generalization of continuous-spin representations and initiated the study of their supersymmetric extensions, establishing a new foundation for subsequent investigations \footnote{For a broad introduction to continuous-spin fields, see the review \cite{BekSk}.
}.

\subsec{Flat:}
In the work by Bekaert and Mourad \cite{l-5-1}, the Wigner equations for CS fields were systematically generalized to Minkowski spacetime of arbitrary dimension. This framework successfully incorporates  Wigner-type equations for mixed-symmetry fields (which naturally arise in higher dimensions) and treats both bosonic and fermionic cases. As already mentioned, a CS field possesses infinitely many spin degrees of freedom. 
 The spectrum of a bosonic CS field encompasses all integer-helicity massless fields, whereas the fermionic counterpart comprises the entire tower of half-integer-helicity fields.
The primary objective of \cite{l-5-1} was to construct an alternative gauge formulation for CS fields.
To this end, building upon the ideas of \cite{l-1-3}, the authors started from a massless spin-$s$ field in $D+1$ dimensions, compactified one dimension to obtain a massive spin-$s$ field in $D$
dimensions within the St\"uckelberg-like formulation, and then took the so-called continuous-spin limit $m\to 0$, $s\to\infty$ while keeping  $s\cdot m = \boldsymbol{\mu}$ fixed. This procedure led to a gauge formulation of CS fields that is analogous to the (Fang-)Fronsdal formulation  \cite{Fronsdal:1978, FangFronsd} for ordinary higher spins. Upon fixing the gauge, they arrived at Wigner-type equations. Following this work, we utilize the fermionic Wigner-like CS equations obtained in this way.

The next surge of interest arose from a series of papers \cite{l-2-1, l-2-2, l-2-3, Schuster:2014hca} investigating the interaction of CS fields with matter. These works constructed a local gauge action -- the Schuster–Toro action -- for bosonic CS fields, which was later generalized to the fermionic case in \cite{l-5-2}, leading to the gauge formulation we mentioned above. Further analysis of the Schuster–Toro action, including its derivation from the Fronsdal-like equations and the study of current exchange between scalar fields via a CS field, was carried out in \cite{l-5-3}. Alternative perspectives on this gauge formulation were also presented in \cite{l-6, l-6-2}.

Further developments of CS theory proceeded along several directions. A BRST analysis of the classical (vector) Wigner CS equations was carried out in \cite{l-1-2} and demonstrated that constructing a Lagrangian within this approach is problematic. This difficulty was overcome in a series of works \cite{takata, 4D-bfi-1, 4D-bfi-3, BIFP-6d, BFI-6D-tf, D6-LC, D6-LD} by developing a spinorial constraint formalism, which enabled explicit BRST Lagrangian descriptions in $4D$ and $6D$ cases. Alternatively, a BRST approach in vector formulation for arbitrary dimensions was developed in \cite{Resh}, where a gauge-invariant Lagrangian for the CS field was constructed using auxiliary fields. 

Mixed-symmetry CS fields were considered in \cite{l-7-1, l-7-2} using Howe duality, which provides a unified description of helicity and continuous-spin fields. Although we do not fully exploit this duality, its algebraic structures are partially reflected in our construction. 

A systematic study of cubic interactions for CS fields was carried out by Metsaev in a series of papers \cite{Me-int-1,Me-int-2,Me-int-3,Me-int-4}. In his most recent work, he employed a Wigner-like covariant formalism, leading to much simpler expressions for the cubic vertices than those obtained in his previous light-cone gauge analysis. Paper \cite{l-6-1} also investigated possible cubic vertices involving a massive (tachyonic) CS field and two massive scalars (see also \cite{l-6-VE} for a discussion of quantization aspects of CS models). In parallel, Schuster, Toro and collaborators explored a wide range of physical implications of CS fields, including interactions with matter, quantum electrodynamics with a CS photon, thermodynamic properties, and possible experimental signatures \cite{l-2-4, Schuster:2023jgc, Schuster:2024wjc, Reilly:2025lnm, Kundu:2025mzm, Kundu:2026zmb}. Also, an on-shell description of CS particles was recently proposed in \cite{Bellazzini:2025}, revealing interesting gravitational effects, including a violation of the equivalence principle at large distances.

Various aspects of supersymmetric extensions of CS models have been addressed in \cite{4D-bfi-2, SCS-N, SCS-NN, SCS-0, SCS-1, l-4-3}. 
\subsec{Curved:} Having reviewed the primary developments in flat spacetime, we now turn to backgrounds of constant curvature, namely $\mathbf{(A)dS}$ spaces. Continuous-spin representations in these spaces have not been extensively studied by physicists, as they do not belong to the standard unitary irreducible representations (UIRs) of the corresponding symmetry groups \footnote{However, such representations have appeared at the group-theoretic level in recent works; see e.g. \cite{JO, KM}.} (see e.g. \cite{BBB, MoldAds}, for a review, see \cite{PL}). While this is indeed the case for de Sitter space, as we elaborate below, the anti-de Sitter ($\mathbf{AdS}$) case offers a significantly more compelling scenario.

A systematic action principle for continuous-spin fields in $\mathbf{(A)dS_D}$\footnote{Notably, this formulation directly yields a description in flat space as well.} was developed by Metsaev \cite{l-3-1}. Following Zinoviev's gauge approach for massive fields \cite{Zinoviev:2001}, he constructed a gauge-invariant Lagrangian involving an infinite tower of symmetric tensor fields. While the standard gauge formulation for the massive case requires the Lagrangian to truncate at a finite spin $s$, Metsaev relaxed this truncation condition, which led to the action depending on two continuous parameters. In anti-de Sitter space, such theories can be unitary for appropriate choices of the parameters, whereas in de Sitter space they inevitably become non-unitary.
Based on this construction, a systematic framework for CS fields in $\mathbf{AdS}$ was established in a series of works  \cite{l-3-2, l-3-8, l-3-5, l-3-6, l-3-7}, encompassing formulations for both bosonic and fermionic mixed-symmetry CS fields. 
The two continuous parameters appearing in these constructions were related to the eigenvalues of the Casimir operators of the $\mathbf{AdS}$ symmetry group, enabling a classification of the continuous-spin UIRs in $\mathbf{AdS}$.

 Furthermore, a frame-like formulation for mixed-symmetry continuous-spin fields in $\mathbf{AdS}$ was developed by Zinoviev \cite{l-4-2}, generalizing Metsaev's earlier metric-like results. For further developments in the same spirit within flat space, see the frame-like analysis in \cite{l-4-1} and the BRST-BV approach in \cite{l-3-9}. A discussion of modified Wigner equations and the connection between the Wigner, Schuster–Toro, and Metsaev formulations can be found in \cite{Najafizadeh:2017tin} (see also work \cite{Htak}, where modified Wigner equations were used to construct unconstrained forms of the free Lagrangian for CS theory in flat spacetime of arbitrary dimension).

The spinorial BRST approach, which proved successful for flat-space CS fields, was extended to $\mathbf{AdS}$ by Buchbinder, Fedoruk, Isaev and Krykhtin \cite{BFIK-4D-ads, BFIK-4D-ads-p, BFIP}. This formulation allowed the construction of a BRST-invariant Lagrangian and established a connection with the representation theory of the $\mathbf{AdS_4}$ symmetry group. For the six-dimensional case, a related analysis, limited to the level of equations of motion and including mixed-symmetry CS fields, was presented in \cite{GP}.

To date, the supersymmetric generalization of CS fields in \(\mathbf{AdS}\) has been addressed only in \cite{SCS-2} for the four-dimensional case.

\subsec{Goals, Novelty and Motivation:}
In this work, we aim to construct Wigner-like equations of motion for symmetric CS fields in $\mathbf{AdS}$ spacetimes of arbitrary dimension for both bosonic and fermionic cases. 
To achieve this, we adopt the framework  established in  \cite{BFIK-4D-ads, BFIK-4D-ads-p}, wherein CS representations in $\mathbf{AdS}$ are constructed via an appropriate deformation of flat-space constraints (equations of motion). 
In contrast to those works mentioned above, we employ a vector (metric-like) formalism rather than a spinorial one. As a starting point, we take the classical $D$-dimensional Wigner equations for bosonic CS fields, and for fermions, the equations introduced in \cite{l-5-1}.
Our formulation ensures that, in the flat-space limit, the equations consistently reduce to the known Wigner-like equations.
Furthermore, utilizing the results of \cite{BFIP, GP}, we demonstrate that the equations obtained through our deformation procedure fix all Casimir operators of the $\mathbf{AdS_D}$ symmetry algebra in the bosonic case, and at least the first two Casimirs in the fermionic case. 
Consequently, these constraints apparently define an irreducible representation of the $\mathbf{AdS_D}$ symmetry algebra. We then analyze the eigenvalues of the Casimir operators obtained on these representations in light of Metsaev's classification of CS UIRs in $\mathbf{AdS}$:
for the bosonic case \cite{l-3-5}, and for the fermionic case in $\mathbf{AdS_4}$ \cite{l-3-7}.

The principal novelty of this work lies in the direct, covariant construction of Wigner-like equations for CS fields in $\mathbf{AdS}$. While a relationship to the gauge formulation of Refs.~\cite{l-3-1, l-3-2} can be anticipated, our direct lifting of the flat-space Wigner equations to the $\mathbf{AdS}$ background has not been previously reported. The simplicity of our constraints is expected to be highly advantageous for a range of future applications.

The motivation for our construction comes from recent work \cite{Me-int-4}, where the Wigner-type equations proved to be a convenient starting point for a systematic and elegant classification of cubic interaction vertices involving continuous-spin fields in flat space. By extending these Wigner-type equations to the $\mathbf{AdS}$ case, we aim to establish a robust foundation for the analysis of cubic vertices involving CS fields in the $\mathbf{AdS}$ background.

Our results also touch upon the holographic aspect of continuous-spin fields. While the work \cite{l-3-5} established a connection between these fields and boundary light-ray operators \cite{Kravchuk:2018htv} in a specific imaginary-parameter regime, our procedure leads to a different sector. Consequently, the fields constructed in this work cannot be directly identified with standard light-ray operators. They may, however, correspond to continuous-spin operators with alternative conformal dimensions, such as those investigated in \cite{SimonCH} -- a possibility that opens an interesting avenue for future holographic analysis.

\subsec{Outline:}
An outline of our paper is as follows. In Section~\ref{Sect: wignereqs}, we briefly review Wigner equations for continuous-spin fields and define our conventions. Section~\ref{Sect: AdSCSequats} proposes a set of constraints (equations of motion) for CS fields in $\mathbf{AdS_D}$ case. In Section~\ref{Sect: Casimirfix}, we demonstrate that the constraints define CS representations with all Casimir operators of the $\mathfrak{so}(2,D-1)$ algebra being fixed; we also discuss the resulting CS representations we obtained within the framework of Metsaev's classification. The fermionic case is developed in a similar manner in Section~\ref{FerCase}. Finally, we present our conclusions in the \hyperref[Discuss]{Discussion}, while auxiliary technical details left for the Appendices.

\section{Wigner continuous-spin equations}\label{Sect: wignereqs}

Continuous-spin representations of the Poincaré group, first introduced by Wigner in his seminal work \cite{w-1}, describe massless particles with an infinite tower of helicities characterized by a real continuous parameter $\boldsymbol{\mu}>0$. In his original $D=4$ construction, Wigner described such representations using relativistic field depending on two vector variables — one for spacetime coordinates and another for an internal "spin" space.

These representations naturally extend to arbitrary spacetime dimensions $D$, where in general they may involve mixed-symmetry tensor fields in their spectrum and require multiple auxiliary variables for their description. However, if one wishes to describe continuous-spin fields using only a single auxiliary vector variable, one is naturally led to restrict attention to the symmetric subclass of continuous-spin representations. This symmetric subclass is characterized by a single continuous parameter $\boldsymbol{\mu}$, just as in the $D=4$ case. It can be described by  functions of type
\be \lb{wave-f}
\Psi(x^m, u^m)\,, \quad \text{where} \;\; \begin{cases}
x^m \;\; \text{are coordinates on} \;\; \mathbb{R}^{1,D-1},\\
u^m \;\; \text{is auxiliary spin} \;\; \mathfrak{so}(1,D-1) \text{-vector}
\end{cases} 
\ee
The Wigner construction for such $D$-dimensional symmetric continuous-spin representations essentially follows the same pattern as in four dimensions. The Poincaré algebra is realized on the space of functions $\Psi(x^m, u^m)$, and the irreducible representation is selected by a set of constraint equations that fix the eigenvalues of the Casimir operators. In this section, we revisit this flat-space construction in detail, as it provides the necessary foundation for our main goal: to generalize the Wigner equations to anti-de Sitter space $\mathbf{AdS_D}$.

To begin, we establish the commutation relations for the $D$-dimensional Poincaré algebra  $\mathfrak{iso}(1,D-1)$:
\bea \lb{s-1-1}
[P_m, P_n] = 0 \,, \;\; [J_{m n}, P_k] = \eta_{n k} P_m - \eta_{m k} P_n \,, \\[7pt]
\lb{s-1-2}
[J_{m n}, J_{k l}] = \eta_{n k} J_{m l} + \eta_{m l} J_{n k} - \eta_{n l} J_{m k} - \eta_{m k} J_{n l} \, , 
\eea
where the indices run as $m,n,k,l = 0, \dots, D-1$ and the Minkowski metric is chosen with the mostly-plus signature
\be \lb{s-3}
\eta_{m n} = \mathrm{diag}(-,\underbrace{+,\dots,+}_{D-1}) \,. 
\ee
This algebra admits a standard realization on the space of functions $\Psi(x^m,u^n)$:
\be\lb{s-4}
P_n = \partial_n \,, \qquad  J_{m n} = \mathcal{L}_{m n} + \mathcal{M}_{m n} \,, 
\ee
where 
\be \lb{s-5}
\mathcal{L}_{m n} := x_m \partial_n - x_n \partial_m \,, \quad \mathcal{M}_{m n} := u_m \frac{\partial}{\partial u^n} - u_n \frac{\partial}{\partial u^m} \,, \qquad \partial_m := \frac{\partial}{\partial x^m}.
\ee

In this realization \eqref{s-4}-\eqref{s-5}, the quadratic and quartic Casimir operators of the Poincar\'e algebra \(\mathfrak{iso}(1,D-1)\) reduce to the following expressions
\be \lb{s-6}
C_2 = - \partial_m \partial^m \,, \quad C_4 = \mathcal{M}_{m}^{\;\;\; k} \mathcal{M}_{n k} \, \partial^m \partial^n + \frac12 \mathcal{M}_{(2)}\,  \partial^k \partial_k  \,,
\ee
where we defined 
\be\lb{s-7}
\mathcal{M}_{(2)} := \mathcal{M}_{m n} \mathcal{M}^{n m} \,. 
\ee 
All higher-order Casimir operators vanish identically within this realization, i.e.
\be \label{higher-Casimirs}
C_{2(\mathfrak{n}+1)} = 0,
\ee
with $n=2, \dots, \lfloor D/2 \rfloor - 1$, which can be verified using the general construction (see, e.g., {\bf Sec. 1.2} of the work \cite{GP}).

First we recall Wigner continuous-spin  equations in a flat spacetime\footnote{See the work \cite{WO-bipf} for a detailed group-theoretical derivation of the most general relativistic equations for the CS field in four dimensions (the classical Wigner CS equations are obtained from them as a special case), including both the vector and spinor formulations.}. Consider the following set of operators acting  on the  functions $\Psi(x^m, u^m)$: 
\bea \lb{s-10}
l_0 := - \partial^m \partial_m \,, \quad l_1:= - i \, \frac{\partial}{\partial u^m} \partial^m \,, \quad 
l_2 := - i \, u^m \partial_m + \boldsymbol{\mu} \,, \quad 
l_3 := \frac{\partial}{\partial u^m} \frac{\partial}{\partial u_m} + 1 \,.
\eea
For a solution $\boldsymbol{\Psi}(x^m, u^n)$ of the Wigner CS equations,
\be \lb{s-10-1}
l_i \boldsymbol{\Psi} = 0, \quad i = 0,\dots,3,
\ee
the Casimir operators of $\mathfrak{iso}(1,D-1)$ take fixed eigenvalues:
\be \lb{s-11}
C_2 \boldsymbol{\Psi} = 0, \qquad 
C_4 \boldsymbol{\Psi} = \boldsymbol{\mu}^2 \boldsymbol{\Psi}.
\ee
We note that the eigenvalues \eqref{s-11} and \eqref{higher-Casimirs}
for the fully symmetric continuous-spin representation are
independent of the particular realization \eqref{s-4}--\eqref{s-5}.
The compatibility of the Wigner equations \eqref{s-10}-\eqref{s-10-1} is ensured by the fact that the operators \(l_{i}\) form a closed operator algebra, whose only non-zero commutation relations are:
\be \lb{s-12}
[l_1,l_2] = l_0 \,, \quad [l_2, l_3] = - 2 l_1.
\ee

Furthermore, the internal space of the auxiliary variables possesses an intrinsic \(\mathfrak{sl}(2)\) symmetry
\footnote{The emergence of this symmetry is a consequence of the oscillator realization of the Lorentz generators and is broadly known as Howe duality (see \cite{Howe} for the original source and \cite{DPC} for a modern review). For applications of this phenomenon in more general but related contexts, see \cite{l-7-1,l-7-2, AGT, AG, AG-1}.}.
 We define the standard set of \(\mathfrak{sl}(2)\) generators:
\be \lb{s-13}
e_0 :=\frac12 (u^m \frac{\partial}{\partial u^m} + \frac{D}{2})\,, \quad e_+ := \frac12 u^m u_m \,, \quad e_- := -\frac12 \frac{\partial}{\partial u^m} \frac{\partial}{\partial u_m} \,,
\ee
which satisfy the $\mathfrak{sl}(2)$ algebra:
\be \lb{s-14}
[e_0, e_{\pm}] = \pm e_{\pm} \,, \quad [e_+,e_-] = 2 e_0.
\ee
The quadratic Casimir operator of this \(\mathfrak{sl}(2)\) algebra is given by
\be \lb{sl2-cas}
\mathfrak{J}^2 = \frac12(e_+ e_- + e_- e_+) + e_0^2 = e_+ e_- + e_0^2 - e_0\,. 
\ee
The Lorentz invariant operator $\mathcal{M}_{(2)}$ can be expressed directly via Casimir operator \eqref{sl2-cas} of the \(\mathfrak{sl}(2)\) algebra as:
\be \lb{s-15}
\mathcal{M}_{(2)} = 8 \, \mathfrak{J}^2 - \frac{D(D-4)}{2}\,. 
\ee

In App.~\ref{appendix:D} we write down the useful commutation relations between the auxiliary vector $u^m$, the derivatives $\partial/\partial u^m$ and \(\mathfrak{sl}(2)\) generators.

\section{$\mathbf{AdS_D}$ Wigner CS equations}\label{Sect: AdSCSequats}

\subsection{Preliminary information}

In this section, we derive a generalization of the Wigner CS equations \eqref{s-10} to the case of $\mathbf{AdS_D}$. By analogy with the flat case, the desired equations will act on the function $\Psi(x^{\mu}, u^m)$, where $x^{\mu}$ are coordinates on $\mathbf{AdS_D}$ and the second argument $u^{m}$ remains an auxiliary spin vector transforming under $\mathfrak{so}(1,D-1)$. 
Throughout this paper, Greek indices $\mu, \nu, \lambda, \dots $  denote world indices, whereas Latin indices  $m,n, k, l, \dots $ represent tangent space ("flat") indices. To treat these two types of indices on an equal footing, we employ the standard vielbein (frame field) formalism.
 Let $g_{\mu \nu}(x)$ be a metric tensor of the $\mathbf{AdS_D}$ space, then the frame fields $e^m_{\mu}$ and $e^{\mu}_m$ are defined via the relations 
$\eta_{m n} e_{\mu}^n e_{\nu}^m = g_{\mu \nu}$ and $g_{\mu \nu} e^{\mu}_{m} e^{\nu}_n = \eta_{m n}$, where $\eta_{m n}$ is the mostly-plus Minkowski metric \eqref{s-3}. 

The covariant derivative acting on the space of functions $\Psi(x^\mu, u^m)$ is defined as
\be\lb{cd-wi}
\mathcal{D}_{\mu} = \partial_{\mu} + \frac12 w_{\mu m n} \mathcal{M}^{m n} \,, \quad  
\ee
where $\partial_{\mu} := \partial/ \partial x^{\mu}$, $w_{\mu m n} = - w_{\mu n m}$ is the spin connection and the operator of spin part of angular momentum $\mathcal{M}_{m n}$ is given in \eqref{s-5}. Moving forward, we will primarily operate with flat indices by introducing the following converted quantities:  
\be \lb{cvd}
e_m := e^{\mu}_m \partial_{\mu} \,, \quad w_{m n k} := e^{\mu}_{m} w_{\mu n k}\, , \quad \mathcal{D}_m := e^{\mu}_m \mathcal{D}_{\mu} = e_m + \frac12 w_{m n k} \mathcal{M}^{n k} \, \quad   \,.
\ee
The commutators of the basis vectors $e_{m}$ and the covariant derivatives $\mathcal{D}_{m}$ satisfy:
\be \lb{cr}
[e_m, e_n] = \mathcal{E}_{m n}^{\;\;\;\;\;l} e_l \,, \quad [\mathcal{D}_m, \mathcal{D}_n] = \mathcal{E}_{m n}^{\;\;\;\;\; k} \mathcal{D}_k +
\frac12 \mathcal{R}_{m n}^{\;\;\;\;\;k l} \mathcal{M}_{k l} \,, 
\ee
where $\mathcal{E}$ is the anholonomy tensor (or the connection free torsion, see e.g. \cite{BK}) and $\mathcal{R}$ is the Riemann curvature tensor. For the anti-de Sitter space, the latter reduces to:
\be \lb{geo-con}
\mathcal{R}_{m n}^{\;\;\;\;\;k l} = - \frac{1}{R^2} \left (\delta_m^k \delta_n^l - \delta_n^k \delta_m^l \right) \,,
\ee
where $R$ is the AdS radius.  The torsion-free condition implies 
\be \lb{geo-con2}
\mathcal{T}_{m n}^{\;\;\;\;\; k} :=\mathcal{E}_{m n}^{\;\;\;\;\; k} + w_{m n}^{\;\;\;\;\; k} - w_{n m}^{\;\;\;\;\; k} = 0 \,.  
\ee
In the subsequent calculations, we will extensively use the connection Laplacian, defined as:
\be \lb{lb}
(\mathcal{D})^2 := \eta^{m n} (\mathcal{D}_m \mathcal{D}_n + w_{m n k} \mathcal{D}^k) \,. 
\ee
along with the following useful commutation relations:
\be \lb{cd-add-v}
[\mathcal{D}_m, u_n] = - w_{m n k} u^k \,, \quad [\mathcal{D}_m, \frac{\partial}{\partial u^n}] = - w_{m n k} \frac{\partial}{\partial u_k} \,.
\ee

Note that in this section and the related Appendix, we do not employ any explicit coordinate charts for $\mathbf{AdS_{D}}$, relying solely on the geometric conditions defined above.

\subsection{Constraints for bosonic CS fields in $\mathbf{AdS_D}$}

To extend the Wigner continuous‑spin equations \eqref{s-10} from flat space $\mathbb{R}^{1,D-1}$ to anti‑de Sitter space $\mathbf{AdS_D}$, we adopt the method developed in works \cite{BFIK-4D-ads-p, BFIK-4D-ads, BFIP, GP}.  
The procedure consists of two main stages:

1.  {\bf Covariantization:}   All ordinary spacetime derivatives  $\partial/\partial x^m$ appearing in the flat-space equations \eqref{s-10} are replaced by the covariant derivatives $\mathcal{D}_m$ defined in \eqref{cvd}.

2. {\bf Deformation of the constraints:} 

To each covariantized constraint we append a compensating term depending on various contractions of the auxiliary spin vector $u^m$ and its derivative $\partial/\partial u^m$.  
Since any such contraction can be rewritten in terms of the $\mathfrak{sl}(2)$ generators \eqref{s-13}, the compensating terms are essentially polynomials in $e_+$, $e_-$ and $e_0$.  

   The compensating term for the first constraint $L_0$ is chosen so that $L_0$ becomes proportional, up to an additive constant, to the quadratic Casimir operator $\mathcal{C}_2$ of the $\mathfrak{so}(2,D-1)$ algebra (see \cite{BFIP,GP}).  
   The remaining deformations $\Delta_a$ ($a=1,2,3$)  are then uniquely determined by requiring that the full set of deformed constraints $\{L_0,L_1,L_2,L_3\}$ forms a closed algebra.  
   By a \textit{closed algebra}
   we mean that the commutator of any two deformed constraints can be expressed as a linear combination of the constraints themselves, where the structure coefficients are allowed to be operators built from $u^m$ and $\partial/\partial u^m$. 

Summarizing this approach, we formulate the following ansatz for the deformed constraints:
\bea
\lb{Lz}
L_0 &=& -\left((\mathcal{D})^2 + \frac{1}{2 R^2} \mathcal{M}_{(2)} + \boldsymbol{\mathfrak{c}}\right)\,, \\[7pt]
\lb{Lo}
L_1 &=& - i \, \frac{\partial}{\partial u^m} \mathcal{D}^m  + \Delta_1\,, \\[7pt]
\lb{Lt}
L_2 &=& - i \, u_m \mathcal{D}^m + \boldsymbol{\mu} + \Delta_2\,, \\[7pt] 
\lb{Ltr}
L_3 &=&  \frac{\partial}{\partial u^m} \frac{\partial}{\partial u_m} + 1 + \Delta_3 \,,
\eea
where we use definitions \eqref{s-7}, \eqref{cvd}, \eqref{lb}, $\boldsymbol{\mathfrak{c}}$  is a constant to be determined, and the deformations  $\Delta_a \, (a=1,2,3)$ are assumed to depend solely on the internal  $\mathfrak{sl}(2)$ generators \eqref{s-13}  and constants. This assumption immediately implies that
\be \lb{OP}
[\mathcal{D}_m, \Delta_a] = 0 \,, \quad a = 1,2,3 \,. 
\ee
By substituting a sufficiently general polynomial ansatz for \(\Delta _{a}\) into the constraints and demanding the algebra of constraints to close, we find that this condition fixes \footnote{The closure condition determines the deformations uniquely up to two constants.  
One corresponds to a shift of the continuous‑spin parameter $\boldsymbol{\mu}$;  
the other is fixed by demanding that the Casimir operators of $\mathfrak{so}(2,D-1)$ become eigenoperators on the representation selected by the constraints.  
Their corresponding eigenvalues will be computed explicitly in the next section.} the coefficients in the ansatz, yielding the following explicit expressions
\be \lb{ans-Del}
\boldsymbol{\mathfrak{c}} =  2 \boldsymbol{\mu} R^{-1} \,, \qquad \Delta_1 = 0 \,, \qquad \Delta_2 = \frac{1}{2 R} \mathcal{M}_{(2)}\,, \qquad \Delta_3 = 0 
\ee

Consequently, having determined the compensating terms $\Delta_a$ and the constant $\boldsymbol{\mathfrak{c}}$, the explicit form of the final constraint operators defining the symmetric continuous-spin field in $\mathbf{AdS_{D}}$ is given by:
\begin{subequations}\label{eq:final-constraints}
\begin{align}
L_0 &= -\left((\mathcal{D})^2 + \frac{1}{2R^2} \mathcal{M}_{(2)} + \frac{2\boldsymbol{\mu}}{R} \right), &
L_1 &= - i \, \frac{\partial}{\partial u^m} \mathcal{D}^m, \\[5pt]
L_2 &= - i \, u_m \mathcal{D}^m + \boldsymbol{\mu} + \frac{1}{2R} \mathcal{M}_{(2)}, &
L_3 &= \frac{\partial}{\partial u^m} \frac{\partial}{\partial u_m} + 1 \,.
\end{align}
\end{subequations}
The resulting commutator algebra of the constraints \eqref{eq:final-constraints} closes as expected. The first non-trivial commutator takes the form: 
\be 
[L_1, L_2] = L_0 + R^{-1}(4 e_0 + 3) L_1 +  4 R^{-1} e_{-} L_2 +  R^{-1} (2 \boldsymbol{\mu} +  R^{-1} \mathcal{M}_{(2)} ) L_3 \,,
\ee
where we utilized \eqref{A-1}, \eqref{A8}  and \eqref{comud-m2}; the second nontrivial commutator reads
\be 
[L_2, L_3] = -2 L_1,
\ee
where we used \eqref{s-15} and \eqref{comUsl}. While all remaining commutators vanish identically:
\be 
[L_0,L_1]=[L_0,L_2]=[L_0,L_3]=0, \,\quad [L_1,L_3]=0 \,.
\ee

\section{CS representation of the $\mathbf{AdS_D}$ symmetry algebra}\label{Sect: Casimirfix}

The isometry algebra of the anti‑de Sitter space $\mathbf{AdS_D}$ is $\mathfrak{so}(2, D-1)$.
In this section, we verify that the constraints $L_i \, (i=0,\dots,3)$ constructed above indeed fix the eigenvalues of the $\mathfrak{so}(2,D-1)$ Casimir operators. Using an explicit realization of the algebra generators in stereographic coordinates, we demonstrate that these constraints are compatible with the $\mathfrak{so}(2,D-1)$ symmetry and compute the eigenvalues of the quadratic and quartic Casimir operators, $\mathcal{C}_2$ and $\mathcal{C}_4$. Finally, we compare our results with Metsaev's classification of unitary irreducible representations of continuous spin.

\subsection{(Isometry) generators and geometric framework}

Let $J_{AB}$ be the generators of the $\mathfrak{so}(2, D-1)$ algebra, where $(A,B = 0,1,\dots D)$. Their commutation relations are given by
\be \lb{JAB}
[J_{AB},J_{CD}] = \eta_{BC}J_{AD} + \eta_{AD}J_{BC} - \eta_{AC}J_{BD} - \eta_{BD}J_{AC} \,, 
\ee
where $\eta_{A B}$ is the embedding space $\mathbb{R}^{2, D-1}$  metric with components $\eta_{AB}=\text{diag}(-,+,\dots,+,-)$.
By splitting the $\mathfrak{so}(2, D-1)$-indices as $A=(m,D)$ with $m=0,\dots,D-1$ we decompose the generators $J_{A B}$ into the $\mathbf{AdS}$ momentum \footnote{We emphasize that, to be consistent with our previous work \cite{GP}, we use the same notation $P_m$ for the $\mathbf{AdS}$ momentum as for the flat momentum; and henceforth, until the end of the paper, by $P_m$ we mean precisely the $\mathbf{AdS}$ case.
}  $P_m$ and the  generators $J_{mn}$ of the Lorentz $\mathfrak{so}(1,D-1)$ subalgebra:
\be \lb{ads-P-J}
P_m := R^{-1} J_{mD},\qquad J_{mn},
\ee
where  $R$ denotes the $\mathbf{AdS}$  radius. The generators $P_m$ and $J_{mn}$ satisfy the algebra
\begin{align}
\label{eq:so-comm-split}
[P_n, P_m] &= R^{-2} J_{nm}, \quad 
[J_{mn}, P_l] = \eta_{nl} P_m - \eta_{ml} P_n, \nonumber \\
[J_{mn}, J_{lk}] &= \eta_{nl} J_{mk} + \eta_{mk} J_{nl} - \eta_{nk} J_{ml} - \eta_{ml} J_{nk} \,, 
\end{align}
which naturally contracts to the Poincaré algebra $\mathfrak{iso}(1,D-1)$ in the flat limit $R\to\infty$ \cite{IW}. 

We realize the $\mathfrak{so}(2, D-1)$ algebra on the same fields $\Psi(x^\mu,u^m)$ introduced in the previous section. These fields depend on the  coordinates $x^\mu$ of the $\mathbf{AdS_D}$ spacetime  and on the auxiliary vector variable $u^m$, which transforms as a flat $\mathfrak{so}(1,D-1)$-vector.

To implement the $\mathbf{AdS_D}$ isometry transformations on $\Psi(x^{\mu}, u^m)$, we employ the Lie–Lorentz derivative (see \cite{YK, WGU} for the original source, \cite{nLLD} for an overview, and \cite{VL,PanKe} for use in modern physics). For a given set of $\mathbf{AdS_{D}}$ Killing vectors  $\phi^\mu_{(\alpha)} \, [\alpha = 1, \dots, D(D+1)/2]$ the Lie–Lorentz derivative is defined as:
\begin{equation}
\lb{LLD}
\mathbb{L}_{\phi_{(\alpha)}} = \phi^\mu_{(\alpha)}\mathcal{D}_\mu 
+ \frac12 \, g_{\lambda\nu}\bigl(\mathcal{D}_\mu\phi^\nu_{(\alpha)}\bigr) e^\mu_a e^\lambda_b\,\mathcal{M}^{ab},
\end{equation}
where in the first term for $\mathcal{D}_{\mu}$ we use \eqref{cd-wi}, in the second term, the covariant derivative should be understood as
\[
\mathcal{D}_{\mu} \phi^{\nu}_{(\alpha)} = \partial_{\mu} \phi^{\nu}_{(\alpha)} + \Gamma^{\nu} _{\mu \rho} \phi^{\rho}_{(\alpha)} \,,
\] 
where $\Gamma_{\mu \rho}^{\nu}$ are Christoffel symbols; and $\mathcal{M}^{ab}$ in \eqref{LLD} is given by (\ref{s-5}). The operator $\mathbb{L}_{\phi_{(\alpha)}}$  
provides a geometrically consistent realization of the spacetime symmetries on the fields $\Psi(x^\mu,u^m)$.

To make further calculations explicit, we adopt stereographic coordinates for $\mathbf{AdS_D}$, in  which the background metric takes a conformally flat form:
\begin{equation}
g_{\mu\nu}(x)=G^{-2}(x)\,\eta_{\mu\nu},\qquad 
G(x)=1-\frac{x^2}{4R^2},\qquad x^2=\eta_{\mu\nu}x^\mu x^\nu, \label{eq:metric}
\end{equation}
with the flat metric $\eta_{\mu\nu}=\text{diag}(-,+,\dots,+)$. In these coordinates, the frame fields (vielbeins) and the spin connection components reduce to:
\begin{align}
e^\mu_n &= G(x) \, \delta^\mu_n,\qquad e_\mu^n = G^{-1}(x) \, \delta_\mu^n, \label{eq:tetrads} \\ \label{eq:spinconn} 
w_{knm} &= -\frac{1}{2R^2}\bigl(x_n\eta_{mk}-x_m\eta_{nk}\bigr), 
\end{align}
where  we define
\begin{align}
x_m  := \eta_{m\mu}x^\mu,\qquad \partial_n := \delta^\mu_n\frac{\partial}{\partial x^\mu}. \label{eq:xpartial}
\end{align}
and  $\delta_m^{\mu}$, $\delta_{\mu}^{m}$ coincide component-wise with $\delta^m_n$, such that $\eta_{m \mu} := \eta_{\nu \mu} \delta^\nu_m$, $\eta^{m \mu} := \eta^{\nu \mu} \delta_\nu^m$.

In these stereographic coordinates (\ref{eq:metric}) the generators \eqref{eq:so-comm-split} of $\mathfrak{so}(2,D-1)$ acquire the explicit form\footnote{The same answer for generators can be obtained using the relation $\mathfrak{so}(2,D-1) \subset \mathfrak{so}(2,D) \simeq \mathfrak{conf}(\mathbb{R}^{1,D-1})$; see, e.g. \cite{IR}.}
\begin{align}
P_n &= e_n - \frac12 w_{nkl}J^{kl}, \label{eq:Pnstereo} \\
J_{mn} &= \mathcal{L}_{mn} + \mathcal{M}_{mn},\qquad 
\mathcal{L}_{mn}=x_m\partial_n-x_n\partial_m \,, \label{eq:Jmnstereo}
\end{align}
where $e_n = e_n^\mu \partial_\mu$.

\subsection{CS representation}

Let us consider  the functions $\boldsymbol{\Psi}(x^{\mu}, u^m)$ for which $L_i \boldsymbol{\Psi} = 0$, where $L_i \, (i=0,1,2,3)$ are defined in \eqref{eq:final-constraints}.
 By utilizing the realization in stereographic coordinates
\eqref{eq:metric} 
one can verify that these constraints \eqref{eq:final-constraints} commute with the  $\mathfrak{so}(2,D-1)$ algebra generators taken in the form \eqref{eq:Pnstereo}-\eqref{eq:Jmnstereo}. Consequently, the equations $L_i\boldsymbol{\Psi} =0$, $i=0,1,2,3$, consistently define a representation of the $\mathfrak{so}(2,D-1)$ algebra on the functions $\boldsymbol{\Psi}(x^{\mu}, u^m)$. To address the irreducibility of this representation, we demonstrate that 
a necessary condition is satisfied: all Casimir operators of the algebra $\mathfrak{so}(2,D-1)$ are fixed on the functions $\boldsymbol{\Psi}(x^{\mu}, u^m)$. 

Accounting for the coordinate realization of the generators  (\ref{eq:Pnstereo})--(\ref{eq:Jmnstereo}) one can represent all Casimir operators of the $\mathfrak{so}(2,D-1)$ algebra in terms of the covariant derivative $\mathcal{D}_n$ and the spin part of the rotation generators $\mathcal{M}_{m n}$. The first two Casimir operators of $\mathfrak{so}(2,D-1)$ are \footnote{These formulas were found in \cite{GP} using the computer algebra system $\mathfrak{Cadabra} \; 2$ \cite{cadKP}.}
\begin{align}
\label{eq:C2-AdS-D}
\mathcal{C}_2 &= - \left(R^2\, (\mathcal{D})^2 + \frac12 \mathcal{M}_{(2)} \right) , \\
\label{eq:C4-AdS-D}
\mathcal{C}_4 &= \, R^2 \left((\mathcal{D}_{_{\bf{(1)}}})^2 + \frac12 \mathcal{M}_{(2)} (\mathcal{D})^2 \right) + \mathcal{O}_4(\mathcal{M}),
\end{align}
where the connection Laplacian $(\mathcal{D})^{2}$ is given by \eqref{lb}, $\mathcal{M}_{(2)}$ by \eqref{s-7} and we recall the explicit definition of the operator $(\mathcal{D}_{\!_{\bf{(1)}}})^{2}$ introduced in \cite{GP} as follows
\be \lb{D1def}
(\mathcal{D}_{\!_{\bf{(1)}}})^2 := \mathcal{M}^{k m} \mathcal{M}_k^{\;\;n} (\mathcal{D}_m \mathcal{D}_n + w_{m n l} \mathcal{D}^l).\,
\ee
The quantity $\mathcal{O}_{4}(\mathcal{M})$ appearing in the quartic Casimir \eqref{eq:C4-AdS-D} is given by
\begin{align} \label{rel-O4-D}
\mathcal{O}_4(\mathcal{M}) &=  \left (\frac{(D-2)(D+1)}{8} \mathcal{M}_{(2)} + \frac18 \mathcal{M}_{(2)}^2  -\frac14 \mathcal{M}_{(4)} \right),
\end{align}
where $\mathcal{M}_{(4)}=\mathcal{M}_{m n}\mathcal{M}^{n k}\mathcal{M}_{k  l}\mathcal{M}^{l m}$. 

It is worth noting that other higher order Casimir operators of $\mathfrak{so}(2,D-1)$ vanish due to the chosen realization for $\mathcal{M}_{mn}$ \eqref{s-5}. Utilizing this realization  $\mathcal{M}_{(4)}$ can be represented in the following form
\be 
\mathcal{M}_{(4)} = 4 \left( \frac{(D-2)(D-3)}{8} \mathcal{M}_{(2)} + \frac{1}{8} \mathcal{M}_{(2)}^2\right).
\ee

We now evaluate these Casimir operators \eqref{eq:C2-AdS-D}-\eqref{eq:C4-AdS-D} on the space of $\boldsymbol{\Psi}(x^\mu, u^m)$ satisfying the constraint system \eqref{eq:final-constraints}.  The eigenvalue of $\mathcal{C}_{2}$ follows immediately from the   $L_{0} \boldsymbol{\Psi} = 0$ equation, where $L_{0}$ is given in \eqref{eq:final-constraints}. 
Evaluating the eigenvalue of $\mathcal{C}_{4}$ requires substituting  \eqref{eq:D1sq-final} into expression \eqref{eq:C4-AdS-D} and again using $L_{0}\boldsymbol{\Psi} = 0$.
 This procedure yields the following  on-shell values of the Casimir operators\footnote{Our eigenvalues are not independent; representations of this type -- continuous degenerate representations -- have already been discussed in the earlier literature \cite{most-1,most-2,most-3,most-4}.}:
\be\label{C2C4eigenb}
\mathcal{C}_2 \boldsymbol{\Psi} = 2 \boldsymbol{\mu} R \, \boldsymbol{\Psi} \,, \qquad \mathcal{C}_4 \boldsymbol{\Psi} =  \boldsymbol{\mu} R \, ( \boldsymbol{\mu} R - (D - 3)) \, \boldsymbol{\Psi}\,.
\ee

In the flat limit $R \to \infty$, 
these expressions correctly contract to the flat Casimir eigenvalues, satisfying $\mathcal{C}_{2,4}/R^2 = C_{2,4}$. The on-shell values \eqref{C2C4eigenb} demonstrate complete agreement\footnote{The agreement is understood taking into account that, as we already discussed before \eqref{ans-Del}, the constraint $L_2$ is defined up to an additive constant.}
 with the results of the papers \cite{BFIP}, \cite{GP}.

To establish the unitarity of the resulting continuous-spin fields, we map our findings to Metsaev's general classification of CS unitary irreducible representations of the $\mathfrak{so}(2,D-1)$ algebra \cite{l-3-5}.  In that framework, representations are labeled by two complex numbers,
 $p$ and $q$, defined via the  equations\footnote{Notice that our convention for the sign of $C_2$ differs from \cite{l-3-5}; we have taken this into account already.}
\be\lb{pqEqbos}
\mathcal{C}_2 = - p^2 - q^2 + \frac{(D-1)^2+(D-3)^2}{4}, \quad
\mathcal{C}_4 = \Bigl(p^2 - \frac{(D-3)^2}{4}\Bigr)\Bigl(q^2 - \frac{(D-3)^2}{4}\Bigr).
\ee

Solving this algebraic system for $p$ and $q$ (up to the interchange $p\leftrightarrow q$) yields a family of solutions that can be written as
\be \lb{pqSolbos}
p = \varepsilon_p \frac12(1+t),\qquad 
q = \varepsilon_q \frac12(1-t),\qquad 
t = \sqrt{(D-2)^2 - 4\boldsymbol{\mu} R},
\ee
where $\varepsilon_p, \varepsilon_q = \pm 1$ represent independent sign choices and recall that $D \geq 4$, $\boldsymbol{\mu} > 0$ and $R > 0$.

\begin{itemize}
\item \textbf{Case (a)}: $\displaystyle \boldsymbol{\mu} > \frac{(D-2)^2}{4R}$,  the parameter $t$ becomes purely imaginary, implying that $p^* = \pm q$. This corresponds to the 
cases \textbf{ii} or \textbf{iii} of the CS UIRs in terminology of \cite{l-3-5}.

\item \textbf{Case (b)}: $ \displaystyle \boldsymbol{\mu} = \frac{(D-2)^2}{4R}$,  \quad with \quad $D>4$, the parameter $t$ is zero $p^2 = q^2$. This corresponds  to  the  case \textbf{vi-c} of the CS UIRs in terminology of \cite{l-3-5}.

\item \textbf{Case (c)}: $\displaystyle \frac{D-3}{R} < \boldsymbol{\mu} < \frac{(D-2)^2}{4R}$, \quad with \quad $D>4$, 
the parameter $t$ is real, $p^2 < x_0$, $q^2 < x_0$, $p^2\neq q^2$, where $x_0 = \left(\frac{D-3}{2}\right)^2$. This corresponds  to  the  case \textbf{vi-a} of the CS UIRs in terminology of \cite{l-3-5}.
\end{itemize}

No other irreducible unitary representations of the continuous-spin type, considered in \cite{l-3-5}, are realized within our framework.

A link between continuous-spin fields and light-ray operators \cite{Kravchuk:2018htv} within the AdS/CFT framework was established in \cite{l-3-5} for the case of purely imaginary parameters $p$ and $q$. In that regime, the conformal dimension ($=(D-1)/2 + p$) and spin ($=(3-D)/2 +q$) of the continuous-spin fields and the boundary operators coincide.
Our analysis demonstrates, however, that such a regime is unattainable for any choice of the parameter $\mathbf{\mu}$ in the present construction; hence, the resulting fields cannot be directly identified with light-ray operators. Nevertheless, continuous-spin operators with alternative conformal dimensions have also been explored in the literature \cite{SimonCH}, and it remains an open question whether the fields constructed in this work correspond to such boundary operators.

\section{Fermionic case}\label{FerCase}

\subsection{Flat space CS constraints}

We now extend our framework to the case of symmetric fermionic continuous-spin fields.
In flat spacetime \cite{l-5-1}, such fields are described by functions 
$\Psi_{\mathrm{A}}(x^m, u^m)$ carrying a Dirac spinor index  $\mathrm{A}$ in $\mathbb{R}^{1,D}$ and its dimension is given by  $2^{\lfloor (D+1)/2 \rfloor}$.  To motivate this higher-dimensional index structure, we recall that a convenient geometric way to construct continuous-spin equations relies on a dimensional reduction procedure. 
Following the approach proposed in \cite{l-5-1}, this formulation originates from a massless spin-$s$ field in $D+1$ dimensions. 
Compactifying one of the dimensions yields a massive spin-$s$ field in $D$ dimensions within a Stückelberg-like framework, from which the continuous-spin limit is subsequently taken by sending $m\to 0$ and $s\to\infty$ while keeping the product $\mu = s \cdot m$ fixed.

Let us introduce the necessary notations.
The higher-dimensional gamma matrices $\Gamma^{\hat M}$ satisfy the Clifford algebra in $(D+1)$ dimensions
\be \label{GamDp1}
\Gamma^{\hat M}\Gamma^{\hat N} + \Gamma^{\hat N}\Gamma^{\hat M} = 2\eta^{\hat M\hat N},
\ee
where $\hat M, \hat N = 0,1,\dots,D-1, D$, $\eta^{\hat M\hat N} = \mathrm{diag}(-,+,\dots,+,+)$.
Splitting the index $\hat M = (m,D)$ 
and denoting the gamma matrix
associated with the extra direction by
\be
\quad \Gamma^{D+1} := \Gamma^{D}
\ee
we obtain
\be \lb{GammaSplit}
\Gamma_m \Gamma_n + \Gamma_n \Gamma_m = 2\eta_{mn},\qquad
\Gamma^{D+1}\Gamma_m + \Gamma_m\Gamma^{D+1}=0,\qquad
(\Gamma^{D+1})^2=1,
\ee
where $m,n=0,1,\dots,D-1$ and $\eta_{mn}$ is the Minkowski metric in $\mathbb{R}^{1,D-1}$.
The matrices $\Gamma_m$ and $\Gamma^{D+1}$ act on the spinor index $\mathrm{A}$ of $\Psi_{\mathrm{A}}(x^m,u^m)$ and are
$
2^{\lfloor (D+1)/2\rfloor}
\times
2^{\lfloor (D+1)/2\rfloor}
$
matrices. Only the algebraic relations \eqref{GammaSplit}, rather than a particular matrix realization, will be essential below. Starting from the $(D+1)$-dimensional Clifford algebra provides a uniform description for both parities of $D$: for even $D$ this reduces to the minimal representation, with $\Gamma^{D+1}$ identified, up to conventional factors, with the usual chirality matrix, whereas for odd $D$ it yields the doubled, non-minimal representation required for the existence of a nontrivial matrix anticommuting with all $\Gamma_m$.

 In flat space,  symmetric fermionic CS representation is realized on the functions $\Psi_{\mathrm{A}}(x^m, u^m)$, which satisfy a set of mixed bosonic and fermionic constraints \cite{l-5-1}
\bea \lb{fl-fer-con}
l_1 &=& - i \, \frac{\partial}{\partial u^m} \partial^m \,, \quad l_2 = - i \, u_m \partial^m + \boldsymbol{\mu} \,, \\
\lb{fl-fer-con-1}
\mathrm{l}_0 &=& - i \, \Gamma_m \partial^m \,, \quad 
\mathrm{l}_3 = \Gamma^m \frac{\partial}{\partial u^m} + \Gamma^{D+1} \,,
\eea
where $\Gamma_{m}$ satisfies the Clifford algebra from \eqref{GammaSplit}.
Note that the fermionic constraints $\mathrm{l}_0$ and $\mathrm{l}_3$ are directly related to the constraints $l_0$ and $l_3$ \eqref{s-10} from Section~\ref{Sect: wignereqs} as follows
\be \lb{fer-bos-flat}
l_0 = \mathrm{l}_0^2 \,, \qquad l_3 = \mathrm{l}_3^2 \,.
\ee

By imposing the constraints \eqref{fl-fer-con}-\eqref{fl-fer-con-1}, we can calculate the eigenvalues of the first three Casimir operators of $\mathfrak{iso}(1,D-1)$:
\be 
C_2 = 0 \,, \qquad C_4 = \boldsymbol{\mu}^2 \,, \qquad C_6 = - \boldsymbol{\mu}^2 \, \frac{(D-3)(D-4)}{8} \,.
\ee
While the eigenvalues of the quadratic and quartic operators $C_2$ and $C_4$ follow immediately  from the constraints, the calculation for $C_6$ utilizes some results of \cite{GP}.

\subsection{$\mathfrak{osp}(1|2)$ algebra}

The intrinsic $\mathfrak{sl}(2)$ symmetry observed in the bosonic internal space is extended in the fermionic sector to an \(\mathfrak{osp}(1\vert{}2)\) superalgebra\footnote{In a more general setting, this symmetry is enhanced to $\mathfrak{osp}(1|2n)$; see, e.g., \cite{l-7-2}.}. 
To demonstrate this, we introduce two fermionic odd operators:
\be 
f_+ = \frac12 \Gamma_m u^m \,, \qquad f_- = \frac12 \Gamma^m \frac{\partial}{\partial u^m} \,. 
\ee
Together with the $\mathfrak{sl}(2)$ generators $e_{\pm}, e_0$ \eqref{s-13}, the generators $f_{\pm}$ form the $\mathfrak{osp}(1|2)$ superalgebra with the relations
\begin{align}
&[e_{+}, f_{-}] = - f_{+}\,,\quad [e_{+}, f_{+}] = 0\,,\quad [e_{-}, f_{+}] = - f_{-}\,, \quad [e_{-}, f_{-}] = 0 \,,  \\
&[e_0, f_{\pm}] = \pm \frac12 f_{\pm}\,, \quad \{f_+, f_-\} = e_0\,, \quad \{f_+, f_+\} = e_+ \,, \quad \{f_-, f_-\} = - e_{-} \,
\end{align}
along with the standard bosonic relations  \eqref{s-14} for the $\mathfrak{sl}(2)$  algebra.

The $\mathfrak{osp}(1|2)$ superCasimir operator is given by:
\be 
\mathbb{J}^2 = \mathfrak{J}^2 - \frac12 [f_+,f_-] \,, 
\ee
where $\mathfrak{J}^2$ was given by \eqref{sl2-cas}.

To properly account for the spinorial degrees of freedom,
we augment the spin part of the Lorentz generators.  Previously, we had defined $\mathcal{M}_{mn} = u_m \frac{\partial}{\partial u^n} - u_n \frac{\partial}{\partial u^m}$; now we consider the following Lorentz generator including the spinor contribution
\be
\mathbb{M}_{mn} := \mathcal{M}_{mn} + \Gamma_{mn}\,,
\ee
where 
\be
\Gamma_{mn} = \frac14[\Gamma_m,\Gamma_n]\,.
\ee
In complete analogy with the bosonic construction \eqref{s-15}, the   Lorentz invariant $\mathbb{M}_{(2)} := \mathbb{M}_{m n} \mathbb{M}^{n m}$  can be expressed through the  $\mathfrak{osp}(1|2)$ superCasimir
\be \lb{suca}
\mathbb{M}_{(2)} = 8 \, \mathbb{J}^2 - \frac{D(D-3)}{4}\,. 
\ee
Furthermore,  taking into account \eqref{fcomMGm2}-\eqref{fcomMGm2ex2}  the fermionic generators satisfy the following anti-commutation relations:
\be \lb{Scas-1}
\left\{f_-,\left( \mathcal{M}_{m n} \Gamma^{n m} + \frac{(D-1)}{2}\right)\right\} = 0,
\ee
\be \lb{Scas-2}
\left\{f_+,\left( \mathcal{M}_{m n} \Gamma^{n m} + \frac{(D-1)}{2}\right)\right\} = 0.
\ee

These relations \eqref{Scas-1}-\eqref{Scas-2} allow us to define the key algebraic object of our construction — the Casimir's ghost operator (or Scasimir) \cite{ABF}

\be \lb{Scas-def}
\mathcal{S} = \left( \mathcal{M}_{m n} \Gamma^{n m} + \frac{(D-1)}{2}\right) = - 2 \, [f_+, f_-] - \frac{1}{2}.
\ee
 The Scasimir operator $\mathcal{S}$ commutes with all bosonic generators of the $\mathfrak{osp}(1|2)$ superalgebra and anti-commutes with the fermionic ones. Its square can be related to the superCasimir  $\mathbb{J}^2$ via:

\be \lb{squareSc}
\mathcal{S}^2 = 4 \mathbb{J}^2 + \frac14\,.
 \ee

\subsection{Fermionic CS constraints in $\mathbf{AdS}$}

We now generalize the flat-space framework to describe fermionic continuous-spin fields in the anti-de Sitter background $\mathbf{AdS_D}$.
These fields in $\mathbf{AdS}$ are realized by functions
\[
\Psi_{\mathrm{A}}(x^{\mu}, u^m),
\]
where, in comparison to the flat-space picture, the flat coordinates
$x^m$ are replaced by
the $\mathbf{AdS}$ coordinates $x^{\mu}$; while the auxiliary spin variable $u^m$ remains unchanged.

Following the same logic as in the bosonic case, we covariantize the flat-space constraints
\eqref{fl-fer-con}-\eqref{fl-fer-con-1} and introduce compensating deformations built from the $\mathfrak{osp}(1|2)$
generators.
Demanding that the resulting constraints form a closed algebra uniquely fixes the deformations.
Throughout this procedure, a key role is played by the Casimir's ghost
(Scasimir) of the $\mathfrak{osp}(1|2)$ algebra, defined in \eqref{Scas-def}.

The final consistent system
 of deformed fermionic continuous-spin constraints in $\mathbf{AdS_{D}}$ takes the following form:
\bea \label{constraintsferm1}
L_1 &=& - i \, \frac{\partial}{\partial u^m} \mathbb{D}^m \,, \quad L_2 = - i \, u_m \mathbb{D}^m + \boldsymbol{\mu} + \frac{1}{2 R} \mathcal{M}_{(2)} + \frac{1}{2 R} \mathcal{M}_{m n} \Gamma^{n m} \,, \\\label{constraintsferm2}\mathbb{L}_0 &=& - i \, \Gamma_m \mathbb{D}^m + \mathbf{\Delta} \,, \quad 
\mathbb{L}_3 = \Gamma^m \frac{\partial}{\partial u^m} + \Gamma^{D+1} \,, 
\eea
where the required deformation term $\mathbf{\Delta}$ is given by 
\be 
\mathbf{\Delta} = - R^{-1} \Gamma^{D+1} \left( \mathcal{M}_{m n} \Gamma^{n m} + \frac{D-1}{2} \right ) = - R^{-1} \Gamma^{D+1} \mathcal{S} \,.
\ee 
Note that in \eqref{constraintsferm1}-\eqref{constraintsferm2}
 the total spin-covariant derivative acting on $\Psi_{\mathrm{A}} (x^{\mu}, u^m)$ is defined as
\be 
\mathbb{D}_m = e_m + \frac12 w_{m n k} \mathbb{M}^{n k} \,.
\ee

The consistency of the system \eqref{constraintsferm1}-\eqref{constraintsferm2} is guaranteed by the closure of the resulting constraint algebra. 
The explicit non-zero commutation and anti-commutation relations evaluate to:
\be \lb{acs-ac} 
\{\mathbb{L}_0, \mathbb{L}_3\} 
=2 L_1 + 2 \mathbf{\Delta} \mathbb{L}_{3}\,, \quad [L_1, \mathbb{L}_0] = R^{-1} \Gamma^{D+1} (L_1 - 2 f_- \mathbb{L}_0) \,,
\ee
\be\lb{fL2fL3}
[L_2, \mathbb{L}_3] = -\mathbb{L}_0 - R^{-1} \mathcal{S} \, \mathbb{L}_3 \,,
\ee
\bea 
\nonumber
[L_{1},L_{2}] &=& \mathbb{L}_0^2 + R^{-1} f_- \mathbb{L}_0 + R^{-1}\left(4 e_0 + \frac{5}{2}\right) L_1 + 4 R^{-1} e_- L_2 \\[7pt] \lb{fL1fL2}
&+& 2 R^{-1}\left( \boldsymbol{\mu} + \frac{1}{2 R} \mathcal{M}_{(2)} + \frac{1}{2 R} \mathcal{M}_{m n} \Gamma^{n m} \right ) \mathbb{L}_3^2 - \frac12 R^{-1} \mathbf{\Delta} \mathbb{L}_3 \,,
\eea
\bea 
\nonumber
[L_{2},\mathbb{L}_{0}] &=& - R^{-1} (2 \Gamma^{D+1} f_+ + 1) \mathbb{L}_0 - 2 R^{-1} f_+ L_1 + R^{-1} \mathbb{L}_3 L_2 \\[7pt] \lb{fL2fL0}
&-& R^{-1}\left( \boldsymbol{\mu} + \frac{1}{2 R} \mathcal{M}_{(2)} + \frac{1}{2 R} \mathcal{M}_{m n} \Gamma^{n m} + R^{-1} \mathcal{S} \right ) \mathbb{L}_3.
\eea
The first anti-commutator in the algebra \eqref{acs-ac} follows from the relation
\bea 
2 \{\mathbf{\Delta}, f_-\} + \{\mathbf{\Delta}, \Gamma^{D+1}\} = 2 \mathbf{\Delta} \mathbb{L}_{3} \,, 
\eea
which is established by employing the anti-commutator \eqref{Scas-1} and the second identity for gamma matrices \eqref{GammaSplit}. 
The second commutator in \eqref{acs-ac} follows from 
the chain of equalities 
\bea \nonumber
[L_1, \mathbb{L}_0] &=& - i [\frac{\partial}{\partial u^m}, \mathbf{\Delta}] \mathbb{D}_{m} + R^{-2} \Gamma_n \frac{\partial}{\partial u^m}\mathbb{M}^{m n} \\[7pt]
\nonumber
&=& R^{-1} \Gamma^{D+1} (L_1 - 2 f_- \mathbb{L}_0) + 2 R^{-1} \Gamma^{D+1} f_- \mathbf{\Delta} + R^{-2} \Gamma_n \frac{\partial}{\partial u^m}\mathbb{M}^{m n} \\[7pt]
&=& R^{-1} \Gamma^{D+1} (L_1 - 2 f_- \mathbb{L}_0) \,,
\eea
where we used \eqref{comUMG}, \eqref{fcomMGm2ex2} and \eqref{Scas-1}. 
The remaining commutators \eqref{fL2fL3}-\eqref{fL2fL0} are calculated in an analogous manner.
As we can see in all calculations involving the algebra of constraints, the properties of Casimir's ghost prove very useful.

Let us briefly summarize how the fermionic constraints are obtained. In constructing the fermionic constraints, we first observe that $[L_1, \mathbb{L}_3] = 0$, so these operators require no deformation. Furthermore, we know that $\mathbb{L}_0$ must be deformed because its square $\mathbb{L}_0^2$ should reproduce the bosonic $L_0$ (with substitution everywhere $\mathcal{M}_{m n} \to \mathbb{M}_{m n}$ and up to a constant). The required deformation $\mathbf{\Delta}$ is then uniquely determined by requiring the anticommutator $\{\mathbb{L}_0, \mathbb{L}_3\}$ to close, yielding the expression presented above. Finally, computing the commutator $[L_1, L_2]$ gives the deformation of $L_2$, which in the fermionic case contains an extra term $\frac{1}{2R} \mathcal{M}_{mn} \Gamma^{nm}$ compared to its bosonic counterpart.

The square of the deformed Dirac operator reads
\be 
\mathbb{L}_0^2 = - ((\mathbb{D})^2 + \frac{1}{R^2} \Gamma_{m n} \mathbb{M}^{n m}) + \{\Gamma_m, \mathbf{\Delta}\} (-i \mathbb{D}^m) + \mathbf{\Delta}^2 \,,
\ee
where the connection Laplacian $(\mathbb{D})^{2}$ is defined as in the bosonic case \eqref{lb}, but with $\mathbb{M}_{mn}$ instead of $\mathcal{M}_{mn}$.
By applying the consequence of the Scasimir squaring identity \eqref{squareSc} 
\be 
R^2 \mathbf{\Delta}^2 = \frac12 \mathbb{M}_{(2)} + \frac{(D-1)(D-2)}{8} \,,
\ee
we  get
\be 
\label{L0square}
\mathbb{L}_0^2 = - \left(\mathbb{D}^2 +\frac{1}{2 R^2} \mathbb{M}_{(2)} + \frac{2 \boldsymbol{\mu}}{R} - \frac{(D-1)(D-2)}{8 R^2}\right) \,.
\ee

\subsection{Eigenvalues of Casimir operators}
Following the bosonic framework  described above, we now construct the $\mathfrak{so}(2,D-1)$ Casimir operators  in the fermionic realization. 
As in the bosonic case, all Casimir operators can be expressed in terms of the covariant 
derivative $\mathbb{D}_n$ and the extended spin part of the rotation generators 
$\mathbb{M}_{mn} = \mathcal{M}_{mn} + \Gamma_{mn}$. 
To ensure clarity and maintain consistency, we explicitly evaluate the first two Casimir operators in the fermionic picture. To distinguish them from their bosonic counterparts, we denote them by $\mathbb{C}_2$ and $\mathbb{C}_4$ 
(instead of $\mathcal{C}_2$ and $\mathcal{C}_4$). They take the form
\begin{align}
\label{eq:C2-AdS-D-ferm}
\mathbb{C}_2 &= - \left(R^2\, (\mathbb{D})^2 + \frac12 \mathbb{M}_{(2)} \right) , \\
\label{eq:C4-AdS-D-ferm}
\mathbb{C}_4 &= \, R^2 \left((\mathbb{D}_{_{\bf{(1)}}})^2 + \frac12 \mathbb{M}_{(2)} (\mathbb{D})^2 \right) + \mathbb{O}_4(\mathbb{M}),
\end{align} 
In analogy with the previous section, the modified Laplacian operator $(\mathbb{D}_{\!_{\bf{(1)}}})^{2}$  is defined as:
\be \label{D1defferm}
(\mathbb{D}_{\!_{\bf{(1)}}})^2 := \mathbb{M}^{k m} \mathbb{M}_k^{\;\;n} (\mathbb{D}_m \mathbb{D}_n + w_{m n l} \mathbb{D}^l).\,
\ee
The quantity $\mathbb{O}_{4}(\mathbb{M})$ in the fermionic case is now given by:
\begin{align} \label{rel-O4-D-ferm}
\mathbb{O}_4(\mathbb{M}) &=  \left (\frac{(D-2)(D+1)}{8} \mathbb{M}_{(2)} + \frac18 \mathbb{M}_{(2)}^2  -\frac14 \mathbb{M}_{(4)} \right),
\end{align}
where $\mathbb{M}_{(4)}=\mathbb{M}_{m n}\mathbb{M}^{n k}\mathbb{M}_{k  l}\mathbb{M}^{l m}$.

We now evaluate the action of these Casimir operators  \eqref{eq:C2-AdS-D-ferm}-\eqref{eq:C4-AdS-D-ferm} on the space of functions \(\mathbf{\Psi}_{\mathrm{A}}(x^\mu, u^m)\) satisfying the fermionic continuous-spin constraints \eqref{constraintsferm1},\eqref{constraintsferm2}. While  the  on-shell value of the quadratic Casimir $\mathbb{C}_{2}$ follows directly from the relation for $\mathbb{L}_{0}^2$ given by \eqref{L0square},  the evaluation of the eigenvalue of the quartic Casimir $\mathbb{C}_{4}$  requires a more involved algebraic reduction, which is partially detailed in App. {\bf \ref{appendix:C}}. This procedure yields the following final eigenvalues for the fermionic continuous-spin fields:
\begin{align}
\mathbb{C}_2 \mathbf{\Psi} &= \left( 2 \boldsymbol{\mu} R-\frac18 (D-1)(D-2)\right) \, \mathbf{\Psi} \,, \\
\lb{c4eigen}
\mathbb{C}_4 \mathbf{\Psi} &=  \left(\left(\boldsymbol{\mu} R - \frac18 (D + 1) (D - 2)\right)^2 + \frac{1}{128}D(D-2)(D-3)(D-5)\right) \, \mathbf{\Psi}\,.
\end{align}

To establish the unitarity of these representations, we analyze our results within the context of Metsaev's classification of unitary irreducible fermionic continuous-spin representations in $\mathbf{AdS_4}$ from \cite{l-3-7}. In this framework, the representations are labeled by the eigenvalues of the Casimir operators in terms of two complex parameters,  $p$ and $q$.

In complete analogy with the bosonic case  \eqref{pqEqbos}-\eqref{pqSolbos} we solve the algebraic system with respect to $p$ and $q$ specifically for $D=4$. This yields:
\be \lb{pqSolFer}
p = \varepsilon_{p} \, \frac{1}{4} \left(1 - \sqrt{25-16 \boldsymbol{\mu} R}\right) \,, \qquad q = \varepsilon_{q}\,  \frac{1}{4} \left(1 + \sqrt{25-16 \boldsymbol{\mu} R}\right) \,,
\ee
where the definition of $\varepsilon_{p,q}$ was given below \eqref{pqSolbos}.  This parameter space reveals three distinct physical regimes for the fermionic continuous-spin fields:

\begin{itemize}
\item \textbf{Case (a) [Principal Series]}: For $\displaystyle \boldsymbol{\mu} R > \frac{25}{16}$ the argument under the radical becomes negative, implying that the parameters are complex conjugate pairs, $p^* = \pm q$. This corresponds to the standard unitary representations of type: $\mathbf{ii\!-\!p^{\frac{1}{2}}}$ or $\mathbf{iii\!-\!p^{\frac{1}{2}}}$.
\item  \textbf{Case (b)  [Exceptional/Limit Series]}: For the critical value $\displaystyle \boldsymbol{\mu} R = \frac{25}{16}$\\
the parameters become real and degenerate, satisfying $p = \pm q$. This bound perfectly reproduces the exceptional unitary representations of type $\mathbf{vi\!-\!c\!-\!2}^{\frac{1}{2} \pm}_k$.
\item \textbf{Case (c) [Complementary and Discrete Series]}: For $ \displaystyle \boldsymbol{\mu} R < \frac{25}{16}$ the parameters $p$ and $q$ are real and non-degenerate ($p \neq \pm q$). Depending on the explicit value of the scale parameter, this domain branches into the following structural subcases: 
\begin{itemize}
    \item \textbf{(c.1) [Complementary Series]}: For $\displaystyle 1 < \boldsymbol{\mu} R < \frac{25}{16}$ with $\boldsymbol{\mu} R\neq3/2$ the configuration realizes the continuous unitary representations of type $\mathbf{vi\!-\!c\!-\!1}^{\frac{1}{2}}_{k}$.
    \item \textbf{(c.2) [Discrete / Anti-Discrete\footnote{To determine whether our representation belongs to the discrete or anti-discrete series, one would need to solve the equations explicitly and examine the helicity expansion of the solution.} Series]}: At the isolated value $\displaystyle \boldsymbol{\mu} R = \frac{3}{2}$ the parameters become $p=0$ and $q=\pm1/2$ (or conversely $q=0$ and $p=\pm1/2$) identifying the discrete series of type
    $\mathbf{vi\!-d}^{\frac{1}{2}}$ or $\mathbf{vi\!-ad}^{\frac{1}{2}}$ ($\mathbf{vi\!-d}^{\frac{1}{2}*}$ or $\mathbf{vi\!-ad}^{\frac{1}{2}*}$)  
    \item \textbf{(c.3)  [Discrete / Anti-Discrete Series Boundary]}: At the lower bound $\boldsymbol{\mu} R = 1$ the model selects the discrete endpoints of type: $\mathbf{vi\!-\!d\!-\!1}^{\frac{1}{2}}_0$ or $\mathbf{vi\!-\!ad\!-\!1}^{\frac{1}{2}}_0$ (with $\varepsilon_q = -1$) and $\mathbf{vi\!-\!d\!-\!1}^{\frac{1}{2}*}_0$ or $\mathbf{vi\!-\!ad\!-\!1}^{\frac{1}{2}*}_0$ (with $\varepsilon_p = -1$).
\end{itemize}
\end{itemize}

\section*{Discussion}
\addcontentsline{toc}{section}{Discussion}
\label{Discuss}
In this work, we have explicitly constructed Wigner-like equations (constraints) for symmetric CS fields in $\mathbf{AdS_D}$ for arbitrary dimensions $D$, covering both the bosonic and fermionic cases. The construction proceeds by first covariantizing the ordinary flat-space derivatives in Wigner constraints and then deforming the resulting operators by adding terms, chosen so that the full set of constraints forms a closed operator algebra. The closure conditions uniquely fix the additional terms and lead to a system of first-class constraints that define a representation of the $\mathfrak{so}(2,D-1)$ isometry algebra. A central role in ensuring the consistency of these constraint algebras is played by specific operators originating from the (super)algebra Howe-dual to the Lorentz subalgebra $\mathfrak{so}(1,D-1)$. 
In the bosonic sector, this corresponds to the Casimir operator of $\mathfrak{sl}(2)$. In the fermionic case, the analogous structure is governed by the Casimir's ghost of the $\mathfrak{osp}(1|2)$ superalgebra. Using the resulting constraint systems, we computed the on-shell values of the quadratic and quartic Casimir operators of $\mathfrak{so}(2,D-1)$ and demonstrated that, for appropriate ranges of the parameter $\boldsymbol{\mu}$, they correspond to certain unitary irreducible representation series classified by Metsaev.

The proposed equations provide a curved-space analogue of the covariant Wigner-type equations recently used for the systematic classification of cubic interactions in flat space \cite{Me-int-4}. We anticipate that our equations will serve as a fertile starting point for an analysis of cubic interactions for continuous-spin fields in the AdS background. A natural direction for future research is to extend the present analysis to mixed-symmetry continuous-spin fields, i.e., to generalise the Bekaert–Mourad equations \cite{l-5-1} to the $\mathbf{AdS}$ background. We also anticipate that our constructions will prove useful for the study of supersymmetric continuous-spin fields in $\mathbf{AdS_D}$, a topic that remains largely unexplored to date.

Since our classification is based on the Casimir operators of free bulk fields, this algebraic structure naturally mirrors the conformal Casimir equations governing two-point functions on the boundary. Although the fields obtained in our construction do not correspond to standard light-ray operators, it remains highly interesting to construct an explicit holographic dictionary for these Wigner-like equations within an alternative sector of boundary CFT operators.

\appendix
\section{Commutator of $L_1$ and $L_2$ }\label{app:CL1L2}
\setcounter{equation}{0}\renewcommand{\theequation}{A.\arabic{equation}}

Consider the commutator of constraints $L_1$ and $L_2$ from \eqref{eq:final-constraints}
\be\lb{A-1}
[L_1, L_2] = - \left[\frac{\partial}{\partial u^m} \mathcal{D}^m, u_n \mathcal{D}^n\right] - i \, \left[\frac{\partial}{\partial u^m}, \frac{1}{2 R} \mathcal{M}_{(2)}\right] \mathcal{D}^m \,,
\ee
where we used \eqref{OP}. 

Let us calculate the first term of \eqref{A-1}
\bea 
\nonumber
\left[\frac{\partial}{\partial u^m} \mathcal{D}^m, u_n \mathcal{D}^n\right] &=& u^n \frac{\partial}{\partial u_m} [\mathcal{D}_{m}, \mathcal{D}_n] + u^n \left [\frac{\partial}{\partial u_m },\mathcal{D}_n \right] \mathcal{D}_m + \left[\frac{\partial}{\partial u_m },u^n\right] \mathcal{D}_m \mathcal{D}_n  \\[7pt]
\lb{lpl-com-ch} 
&+& \frac{\partial}{\partial u_m } [\mathcal{D}_m, u^n] \mathcal{D}_n \,. 
\eea

Let us now find all commutators from \eqref{lpl-com-ch}  separately. The first commutator reads
\bea \nonumber
u^n \frac{\partial}{\partial u_m} [\mathcal{D}_m, \mathcal{D}_n] &=& u^n \frac{\partial}{\partial u_m} \left(\mathcal{E}_{m n}^{\;\;\;\;\; l} \mathcal{D}_l + \frac12\mathcal{R}_{m n}^{\;\;\;\;\;k l} \mathcal{M}_{k l} \right) \\[7pt] 
\lb{A2}
&=& u^n \frac{\partial}{\partial u_m} \mathcal{E}_{m n}^{\;\;\;\;\; l} \mathcal{D}_l - \frac{1}{R^2} u^n \frac{\partial}{\partial u_m} \mathcal{M}_{m n} \,,
\eea
where we used the second eq. \eqref{cr} and the first from \eqref{geo-con}. 
The second one from \eqref{lpl-com-ch} can be represented as 
\be \lb{A3}
u^n[\frac{\partial}{\partial u_m},\mathcal{D}_n] \mathcal{D}_m = - u^n \frac{\partial}{\partial u_m} w_{n m l} \mathcal{D}^{l}\,,
\ee
where we used \eqref{cd-add-v}
The third term from \eqref{lpl-com-ch} is
\be \lb{A4}
\left[\frac{\partial}{\partial u_m },u^n\right] \mathcal{D}_m \mathcal{D}_n = \mathcal{D}_n \mathcal{D}^n \,.
\ee
The last commutator comes to
\be \lb{A5}
\frac{\partial}{\partial u_m } [\mathcal{D}_m, u^n] \mathcal{D}_n = \frac{\partial}{\partial u_m } u^n w_{m n l} \mathcal{D}^l \, ,
\ee
where we used \eqref{cd-add-v}. 
By now combining the first term from \eqref{A2} and \eqref{A3}, we obtain 
\be \lb{A6}
u^n \frac{\partial}{\partial u_m} (\mathcal{E}_{m n}^{\;\;\;\;\; l}  - w_{n m l} ) \mathcal{D}_l = - u^n \frac{\partial}{\partial u_m} w_{m n l} \mathcal{D}^l \,,
\ee
where the second eq. from \eqref{geo-con} are taken into account. Combining \eqref{A6} with \eqref{A5} we get
\be \lb{A7}
\left( \frac{\partial}{\partial u_m } u^n - u^n \frac{\partial}{\partial u_m}\right) w_{m n l} \mathcal{D}^l = w_{n}^{\;\; n l} \mathcal{D}_l\,.
\ee
Now we have to sum the second term of \eqref{A2}, \eqref{A4} and \eqref{A7}, so we get
\be \lb{A8}
\left[\frac{\partial}{\partial u^m} \mathcal{D}^m, u_n \mathcal{D}^n\right] =  (\mathcal{D})^2 - \frac{1}{R^2} u^n \frac{\partial}{\partial u_m} \mathcal{M}_{m n} = (\mathcal{D})^2 - \frac{1}{2 R^2} \mathcal{M}_{(2)} \,,
\ee
where we used the definition of the operator $(\mathcal{D})^2$ given by \eqref{lb}.

\section{Action of the $(\mathcal{D}_{(1)})^2$ operator on  $\boldsymbol{\Psi}(x^{\mu}, u^m)$}\label{app:D(1)-action}
\setcounter{equation}{0}\renewcommand{\theequation}{B.\arabic{equation}}

In this appendix, we compute the action of the operator $(\mathcal{D}_{(1)})^2$ on fields $\boldsymbol{\Psi}(x^\mu, u^m)$ satisfying the constraint equations $L_i \boldsymbol{\Psi} = 0$ ($i = 0,1,2,3$), where $L_i$ have written in \eqref{eq:final-constraints}. This computation is necessary for determining the eigenvalues of the second Casimir operator of the $\mathfrak{so}(2,D-1)$ algebra on functions $\boldsymbol{\Psi}(x^{\mu}, u^m)$.

\paragraph{Transformation of the $(\mathcal{D}_{(1)})^2$ operator}
Using explicit representation for $\mathcal{M}_{m n}$ from \eqref{s-5} in the definition \eqref{D1def} of $(\mathcal{D}_{(1)})^2$ we get
\bea \nonumber
(\mathcal{D}_{(1)})^2 &=& 2 \left(e_{+} \left(\frac{\partial}{\partial u} \cdot \mathcal{D}\right)^2 -   e_{-} (u \cdot \mathcal{D})^2 -  e_0 \left\{(u \cdot \mathcal{D}), \left(\frac{\partial}{\partial u} \cdot \mathcal{D}\right)\right\}\right) \\[7pt]
\lb{Dtra}
&+& \frac{(D-4)}{2}(\mathcal{D})^2  - \frac{(D-2)}{4 R^2} \mathcal{M}_{(2)} \,,
\eea
where $\{\,,\}$ means anticommutator, and we use the shorthand notation $X\cdot Y := X^m Y_m$ for Lorentz contractions.  Or using \eqref{A8} it can be represented in form 
\bea \nonumber
(\mathcal{D}_{(1)})^2 &=& 2 \left(e_{+} \left(\frac{\partial}{\partial u} \cdot \mathcal{D}\right)^2 -   e_{-} (u \cdot \mathcal{D})^2 -  2 e_0 (u \cdot \mathcal{D}) \left(\frac{\partial}{\partial u} \cdot \mathcal{D}\right)\right) \\[7pt]
\lb{Dtra-1}
&+& \left(\frac{(D-4)}{2}-2 e_0 \right)(\mathcal{D})^2   - \left(\frac{(D-2)}{4 R^2} -\frac{1}{R^2} e_0 \right) \mathcal{M}_{(2)} \,,
\eea

\paragraph{Consequences of the constraints}

From the definitions \eqref{eq:final-constraints} of the operators $L_i$ and the conditions $L_i \boldsymbol{\Psi} = 0$, we immediately derive:

\bea \label{eq:C-constraints-0}
(\mathcal{D})^2 \boldsymbol{\Psi} &=& \left( - \frac{2 \boldsymbol{\mu}}{R} - \frac{1}{2 R^2}\mathcal{M}_{(2)} \right) \boldsymbol{\Psi}, \\[7pt]
\label{eq:C-constraints-1}
\left(\frac{\partial}{\partial u} \cdot \mathcal{D}\right) \boldsymbol{\Psi} &=& 0, \\[7pt]
\label{eq:C-constraints-2}
(u \cdot \mathcal{D}) \boldsymbol{\Psi} &=& - i \left( \boldsymbol{\mu} + \frac{1}{2R} \mathcal{M}_{(2)} \right) \boldsymbol{\Psi}, \\[7pt]
\label{eq:C-constraints-3}
e_{-} \boldsymbol{\Psi} &=& \frac12 \boldsymbol{\Psi},
\eea
where we used definition \eqref{s-13} for $e_-$ .

\paragraph{Computation of $(u \cdot \mathcal{D})^2 \boldsymbol{\Psi}$ and related quantities}

From \eqref{eq:C-constraints-1}-\eqref{eq:C-constraints-2} and the first commutator form \eqref{comud-m2} we find:

\bea \label{eq:uDuD2}
(u \cdot \mathcal{D})^2 \boldsymbol{\Psi} &=& \left[- \left( \boldsymbol{\mu} + \frac{1}{2R} \mathcal{M}_{(2)} \right)^2 
- \frac{1}{2R} \left( \boldsymbol{\mu} + \frac{1}{2R} \mathcal{M}_{(2)} \right) \left( -8 e_0 + 6 \right) \right] \boldsymbol{\Psi}.
\eea

To compute $e_{-} (u \cdot \mathcal{D})^2 \boldsymbol{\Psi}$, we note that only the term proportional to $e_0$ in \eqref{eq:uDuD2} does not commute with $e_{-}$. Using $[e_{-}, e_0] = e_{-}$ and $e_{-} \boldsymbol{\Psi} = \frac12 \boldsymbol{\Psi}$, we obtain after a short calculation:

\be \label{eq:e-uD2-result}
e_{-} (u \cdot \mathcal{D})^2 \boldsymbol{\Psi} = \frac12 (u \cdot \mathcal{D})^2 \boldsymbol{\Psi} + \frac{2}{R} \left( \boldsymbol{\mu} + \frac{1}{2R} \mathcal{M}_{(2)} \right) \boldsymbol{\Psi}.
\ee

\paragraph{Evaluation of $(\mathcal{D}_{(1)})^2 \boldsymbol{\Psi}$}

Using expression \eqref{Dtra-1} for $(\mathcal{D}_{(1)})^2$ and noting that due to \eqref{eq:C-constraints-1} terms containing $\left( \frac{\partial}{\partial u} \cdot \mathcal{D} \right)$ vanish on the $\boldsymbol{\Psi}$, we have:

\bea \label{eq:D1sq-intermediate}
(\mathcal{D}_{(1)})^2 \boldsymbol{\Psi} &=& \Bigl[ -2 e_{-} (u \cdot \mathcal{D})^2 + \left( \frac{(D-4)}{2} - 2 e_0 \right) (\mathcal{D})^2 \nn \\
&& - \left( \frac{(D-2)}{4 R^2} - \frac{1}{R^2} e_0 \right) \mathcal{M}_{(2)} \Bigr] \boldsymbol{\Psi}.
\eea

Substituting \eqref{eq:C-constraints-0} and \eqref{eq:e-uD2-result} into \eqref{eq:D1sq-intermediate}, and simplifying, we observe that all terms containing $e_0$ cancel mutually.

After this cancellation, we obtain the final result:

\be \label{eq:D1sq-final}
(\mathcal{D}_{(1)})^2 \boldsymbol{\Psi} = \left( \frac{(2-D+2\boldsymbol{\mu} R)}{2 R^2} \mathcal{M}_{(2)} + \frac{1}{ 4 R^2} \mathcal{M}_{(2)}^2+\boldsymbol{\mu} \left( \boldsymbol{\mu} - \frac{D-3}{R} \right) \right) \boldsymbol{\Psi} \, .
\ee

\section{Simplification of the $(\mathbb{D}_{(1)})^2$ operator}
\label{appendix:C}
\setcounter{equation}{0}\renewcommand{\theequation}{C.\arabic{equation}}

To calculate the quartic Casimir $\mathbb{C}_{4}$ \eqref{eq:C4-AdS-D-ferm} in the fermionic case we first compute its leading contribution,
$R^2(\mathbb{D}_{(1)})^2$ \eqref{D1defferm}. To this end, we decompose it into symmetric and antisymmetric parts with respect to the indices $m,n$ as \footnote{Throughout the appendices, parentheses $(m n)$ denote symmetrization and brackets $[m n]$ denote antisymmetrization, both with factor $1/2$: $A_{(m}B_{n)} = \frac12(A_mB_n + A_nB_m)$, $A_{[m}B_{n]} = \frac12(A_mB_n - A_nB_m)$.}
\bea 
\nonumber
R^2(\mathbb{D}_{(1)})^2 &=& -R^2 (\mathbb{M}_{k (m} \mathbb{M}_{n)}^{\;\;\;k} + \mathbb{M}_{k [m} \mathbb{M}_{n]}^{\;\;\;k}) (\mathbb{D}^{m} \mathbb{D}^n + w^{m n l} \mathbb{D}_l) \\[7pt] \label{D1appsymantisym}
&=& - R^2\mathbb{M}_{k (m} \mathbb{M}_{n)}^{\;\;\;k}(\mathbb{D}^{m} \mathbb{D}^n + w^{m n l} \mathbb{D}_l) - \frac{D-2}{4} \mathbb{M}_{(2)} \,, 
\eea
where for the antisymmetric part we have utilized the following algebraic identity 
$$ 
\mathbb{M}_{km}\mathbb{M}_{\;\;n}^{k}\mathbb{M}^{mn}=\left(\frac{D-2}{2}\mathbb{M}_{mn}+\frac{D-1}{2}\eta_{mn}\right)\mathbb{M}^{mn}=-\frac{D-2}{2}\mathbb{M}_{(2)}.
$$
Next, we expand the symmetric product of the full generators $\mathbb{M}_{m n}$, separating the purely bosonic part $\mathcal{M}_{mn}$ and the remaining terms
\bea \nonumber
\mathbb{M}_{k (m} \mathbb{M}_{n)}^{\;\;\; k} &=& \mathcal{M}_{k (m} \mathcal{M}_{n)}^{\;\;\; k}  + \frac{D-5}{4}\eta_{m n}+ \\[7pt] \lb{symp2Mbb}
 &+&  f_- (u_m \Gamma_n + u_n \Gamma_m) -  f_+ \left(\frac{\partial}{\partial u^m} \Gamma_n + \frac{\partial}{\partial u^n} \Gamma_m\right)\,,
\eea
where we used \eqref{GGsym}. By substituting the explicit form of the purely bosonic contribution, expressed in terms of the auxiliary variables $u^{m}$ and $\partial/\partial u^m$: 

\bea \nonumber
\mathcal{M}_{k (m} \mathcal{M}_{n)}^{\;\;\; k} &=& 2 e_- u_m u_n - 2 e_+ \frac{\partial}{\partial u^m} \frac{\partial}{\partial u^n} \\[7pt]
&+& \left(2 e_0 - \frac{(D-4)}{2}\right) \eta_{m n} + 2 e_0 \left(u_m \frac{\partial}{\partial u^n} + u_n \frac{\partial}{\partial u^m}\right), 
\eea
the first term on the right-hand side of the relation \eqref{D1appsymantisym} can be cast into a following sum:
\be \lb{same-md}
- \mathbb{M}_{k (m} \mathbb{M}_{n)}^{\;\;\;k}(\mathbb{D}^{m} \mathbb{D}^n + w^{m n l} \mathbb{D}_l) = \mathbb{X} + \mathbb{Y}\,, 
\ee
where these operators $\mathbb{X}$ and $\mathbb{Y}$ are defined as follows:
\bea \nonumber
\mathbb{X} &=& - \mathcal{M}_{k (m} \mathcal{M}_{n)}^{\;\;\; k} (\mathbb{D}^{m} \mathbb{D}^n + w^{m n l} \mathbb{D}_l)  \\[7pt] \nonumber
&=& 2 \left(e_{+} \left(\frac{\partial}{\partial u} \cdot \mathbb{D}\right)^2 -   e_{-} (u \cdot \mathbb{D})^2 -  2 e_0 (u \cdot \mathbb{D}) \left(\frac{\partial}{\partial u} \cdot \mathbb{D}\right)\right) \\[7pt]
\lb{Dtran-1}
&+& \left(\frac{(D-4)}{2}-2 e_0 \right)(\mathbb{D})^2  +\frac{2}{R^2} e_0 \, u_n \frac{\partial}{\partial u^m} \mathbb{M}^{m n}
\eea
and 
\bea \nonumber
\mathbb{Y} &=& - \frac{(D-5)}{4} (\mathbb{D})^2 - f_- (u_m \Gamma_n + u_n \Gamma_m) (\mathbb{D}^{m} \mathbb{D}^n + w^{m n l} \mathbb{D}_l) \\[7pt] \nonumber
&+& f_+\left(\frac{\partial}{\partial u^m} \Gamma_n + \frac{\partial}{\partial u^n} \Gamma_m\right)(\mathbb{D}^{m} \mathbb{D}^n + w^{m n l} \mathbb{D}_l) \\[7pt] \nonumber
&=& - \frac{(D-5)}{4} (\mathbb{D})^2 - f_- \left( 2 (\Gamma \cdot \mathbb{D})(u\cdot \mathbb{D}) - \frac{1}{R^2} \Gamma_n u_m \mathbb{M}^{m n} \right) \\[7pt] \lb{Dtran-2}
&+&f_+ \left(2 (\Gamma \cdot \mathbb{D})\left(\frac{\partial}{\partial u} \cdot \mathbb{D}\right) - \frac{1}{R^2} \Gamma_n \frac{\partial}{\partial u^m} \mathbb{M}^{m n}\right).  
\eea
We recall the shorthand notation $X\cdot Y := X^m Y_m$ for Lorentz contractions introduced in Appendix {\bf \ref{app:D(1)-action}}.

Then, substituting \eqref{same-md} together with \eqref{Dtran-1}--\eqref{Dtran-2} into \eqref{D1appsymantisym}, we obtain a new representation of $(\mathbb{D}_{(1)})^2$ within our framework. 
Using this result in the definition of the quartic Casimir \eqref{eq:C4-AdS-D-ferm} allows us to directly employ the fermionic constraints \eqref{constraintsferm1}--\eqref{constraintsferm2}.
Then, performing algebraic manipulations based on relations of Appendix \ref{appendix:D} and the identity 
\bea
\nonumber
\mathbb{M}_{(4)} 
&=& \mathbb{M}^{(m}_{\;\;\;k} \mathbb{M}^{n) k} \mathbb{M}_{r (n} \mathbb{M}^{r}_{\;\;m)}
+ \frac{(D-2)^2}{4} \mathbb{M}_{(2)} \,=\\ 
 &=& -\frac{1}{32} D (D-1) (D-2) (D-3) + \frac{1}{4} (D - 3 ) ( D -2 ) \, \mathbb{M}_{(2)} + \frac{1}{2} \mathbb{M}_{(2)}^2 \,,
\eea
we obtain the eigenvalue \eqref{c4eigen} of the quartic Casimir $\mathbb{C}_4$.

\section{Useful identities and commutation relations}
\label{appendix:D}

\setcounter{equation}{0}\renewcommand{\theequation}{D.\arabic{equation}}

For practical calculations and further generalization to $\mathbf{AdS}$ spacetime, the following commutation relations between the auxiliary vector, its derivatives and \(\mathfrak{sl}(2)\) generators are highly useful:
\be 
\lb{comUsl}
[u_m, e_+] = 0\,, \quad [u_m, e_-] = \frac{\partial}{\partial u^m} \,, \quad
[u_m, e_0] = - \frac12 u_m \,. 
\ee

\be
\lb{comDUsl}
[\frac{\partial}{\partial u^m}, e_-] = 0\,, \quad [\frac{\partial}{\partial u^m}, e_+] = u_m \,, \quad [\frac{\partial}{\partial u^m}, e_0] = \frac12 \frac{\partial}{\partial u^m} \,. 
\ee
The commutators of the auxiliary vector, its derivatives with the operator \(\mathcal{M}_{(2)}\) take the form:
\be 
\lb{comud-m2}
[u_m, \mathcal{M}_{(2)}] = -8 e_0 u_m + 6 u_m + 8 e_{+} \frac{\partial}{\partial u^m}\,, \quad [\frac{\partial}{\partial u^m},\mathcal{M}_{(2)}] = 8 e_0 \frac{\partial}{\partial u_m} + 6 \frac{\partial}{\partial u_m} + 8e_{-} u_m\,.
\ee
For various contractions of two gamma-matrices we have 
\bea \nonumber
\Gamma_{m n} \Gamma^{n}_{\;\;k} &=& \frac{(D-1)}{4} \eta_{m k} + \frac{(D-2)}{2} \Gamma_{m k}\,, \qquad
\Gamma_{k (m} \Gamma_{n)}^{\;\;\; k} = \frac{(D-1)}{4} \eta_{m n}\,, \\[7pt] \lb{GGsym}
\Gamma_{m n} \Gamma^{n m} &=& \frac{D(D-1)}{4}
\eea
For contraction of $\mathcal{M}_{m n}$ and $\Gamma_{m n}$ in terms of $\mathfrak{osp}(1|2)$ generators we get
\bea \lb{MGcontr}
\nonumber
\mathcal{M}_{m n} \Gamma^{n m} &=& 2 \, e_0 - 4 \, f_+ f_- - \frac{D}{2} = 2(f_+ f_- + f_- f_+) - 4 f_+ f_- - \frac{D}{2} \\[7pt]
&=& -2 (f_+ f_- - f_- f_+) - \frac{D}{2}
\eea
We also need the following formulae
\bea 
\nonumber
[f_-, \mathcal{M}_{m n} \Gamma^{n m}] &=& - \Gamma_n \frac{\partial}{\partial u^{m}} \mathbb{M}^{m n} \,, \quad [f_+,\mathcal{M}_{m n} \Gamma^{n m}] = - \Gamma_n u_m \mathbb{M}^{m n} \,, \\[7pt] \lb{fcomMGm2} [f_-, \mathcal{M}_{(2)}] &=& 2 \Gamma_n \frac{\partial}{\partial u^{m}} \mathbb{M}^{m n}\,,
\eea
where R.H.S's also can be represented in the form
\be \lb{fcomMGm2ex1}
\Gamma_n u_m \mathbb{M}^{m n} = f_+ - 4 f_+ e_0 + 4 e_+ f_- = -2 f_+ \left(\mathcal{M}_{m n} \Gamma^{n m} + \frac{(D-1)}{2}\right) \,, 
\ee
\be \lb{fcomMGm2ex2}
 \Gamma_n \frac{\partial}{\partial u^m} \mathbb{M}^{m n} = - f_- + 4 e_0 f_- + 4 f_+ e_- = 2 \left(\mathcal{M}_{m n} \Gamma^{n m} + \frac{(D-1)}{2}\right) f_{-}\,.
\ee
Similar to the formulae \eqref{comUsl}-\eqref{comDUsl} let us also find
\bea
\lb{comUMG}
[\frac{\partial}{\partial u^m}, \mathcal{M}_{n l} \Gamma^{l n}] &=& -\frac{\partial}{\partial u^m} + 2 f_- \Gamma_m \,, \quad [u_m,\mathcal{M}_{n l} \Gamma^{l n}] = - u_m + 2 f_+ \Gamma_m \\[7pt]
\lb{comGMG}
[\Gamma_m, \mathcal{M}_{n l} \Gamma^{l n}] &=& 2 \Gamma_m - 4 f_- u_m + 4 f_+ \frac{\partial}{\partial u^m} \,. 
\eea

\end{document}